\documentclass[journal]{vgtc}                % final (journal style)
\ifpdf%                                % if we use pdflatex
  \pdfoutput=1\relax                   % create PDFs from pdfLaTeX
  \usepackage{graphicx}                % allow us to embed graphics files
  \DeclareGraphicsExtensions{.pdf,.png,.jpg,.jpeg} % for pdflatex we expect .pdf, .png, or .jpg files
\else%                                 % else we use pure latex
  \usepackage{graphicx}                % allow us to embed graphics files
  \DeclareGraphicsExtensions{.eps}     % for pure latex we expect eps files
\fi%

\graphicspath{{figures/}{pictures/}{images/}{./}{./IEEEVR_2018_JOURNAL/}} % where to search for the images

\usepackage{microtype}                 % use micro-typography (slightly more compact, better to read)
\PassOptionsToPackage{warn}{textcomp}  % to address font issues with \textrightarrow
\usepackage{textcomp}                  % use better special symbols
\usepackage{mathptmx}                  % use matching math font
\usepackage{times}                     % we use Times as the main font
\usepackage{cite}                      % needed to automatically sort the references
\usepackage{tabu}                      % only used for the table example
\usepackage{booktabs}                  % only used for the table example
\usepackage{color}
\usepackage{comment}
\usepackage{subfigure}
\usepackage{amssymb}

\usepackage{enumitem}
\def\ie{i.e.\ }
\def\eg{e.g.\ }
\def\etal{et~al.\ }
\usepackage[normalem]{ulem}

\usepackage{tabularx}
\newcolumntype{L}[1]{>{\raggedright\arraybackslash}m{#1}}
\newcolumntype{C}[1]{>{\centering\arraybackslash}m{#1}}
\newcolumntype{R}[1]{>{\raggedleft\arraybackslash}m{#1}}

\usepackage{algorithm}
\usepackage{algpseudocode}
\usepackage{xspace}
\algrenewcommand\algorithmicrequire{\textbf{Input:}}
\algrenewcommand\algorithmicensure{\textbf{Output:}}
\algnewcommand\And{\textbf{and}\xspace}
\algnewcommand\Or{\textbf{or}\xspace}
\algnewcommand\Not{\textbf{not}\xspace}

\renewcommand\copyrighttext{\begin{minipage}[c]{\textwidth}Copyright (c) 2019 IEEE. Personal use is permitted. For any other purposes, permission must be obtained from the IEEE by emailing pubs-permissions@ieee.org. This is the author's version of an article that has been published in this journal. The final version of record is available at \url{http://dx.doi.org/10.1109/TVCG.2019.2899231}.\end{minipage}}

\onlineid{1254}

\vgtccategory{Research}
\vgtcpapertype{system}

\title{SLAMCast: Large-Scale, Real-Time 3D Reconstruction and Streaming for Immersive Multi-Client Live Telepresence}

\author{Patrick Stotko, Stefan Krumpen, Matthias B. Hullin, Michael Weinmann, and Reinhard Klein}
\authorfooter{
\item
 Patrick Stotko, Stefan Krumpen, Matthias B. Hullin, Michael Weinmann, and Reinhard Klein are with University of Bonn. E-mail: \{stotko, krumpen, hullin, mw, rk\}@cs.uni-bonn.de.
}

\shortauthortitle{Stotko \MakeLowercase{\textit{et al.}}: SLAMCast: Large-Scale, Real-Time 3D Reconstruction and Streaming for Immersive Multi-Client Live Telepresence}
\abstract{
Real-time 3D scene reconstruction from RGB-D sensor data, as well as the exploration of such data in VR/AR settings, has seen tremendous progress in recent years.
The combination of both these components into telepresence systems, however, comes with significant technical challenges.
All approaches proposed so far are extremely demanding on input and output devices, compute resources and transmission bandwidth, and they do not reach the level of immediacy required for applications such as remote collaboration.
Here, we introduce what we believe is the first practical client-server system for real-time capture and many-user exploration of static 3D scenes.
Our system is based on the observation that interactive frame rates are sufficient for capturing and reconstruction, and real-time performance is only required on the client site to achieve lag-free view updates when rendering the 3D model.
Starting from this insight, we extend previous voxel block hashing frameworks by introducing a novel thread-safe GPU hash map data structure that is robust under massively concurrent retrieval, insertion and removal of entries on a thread level.
We further propose a novel transmission scheme for volume data that is specifically targeted to Marching Cubes geometry reconstruction and enables a 90\% reduction in bandwidth between server and exploration clients.
The resulting system poses very moderate requirements on network bandwidth, latency and client-side computation, which enables it to rely entirely on consumer-grade hardware, including mobile devices.
We demonstrate that our technique achieves state-of-the-art representation accuracy while providing, for any number of clients, an immersive and fluid lag-free viewing experience even during network outages.
} % end of abstract


\teaser{
    \centering
    \includegraphics[width=\textwidth]{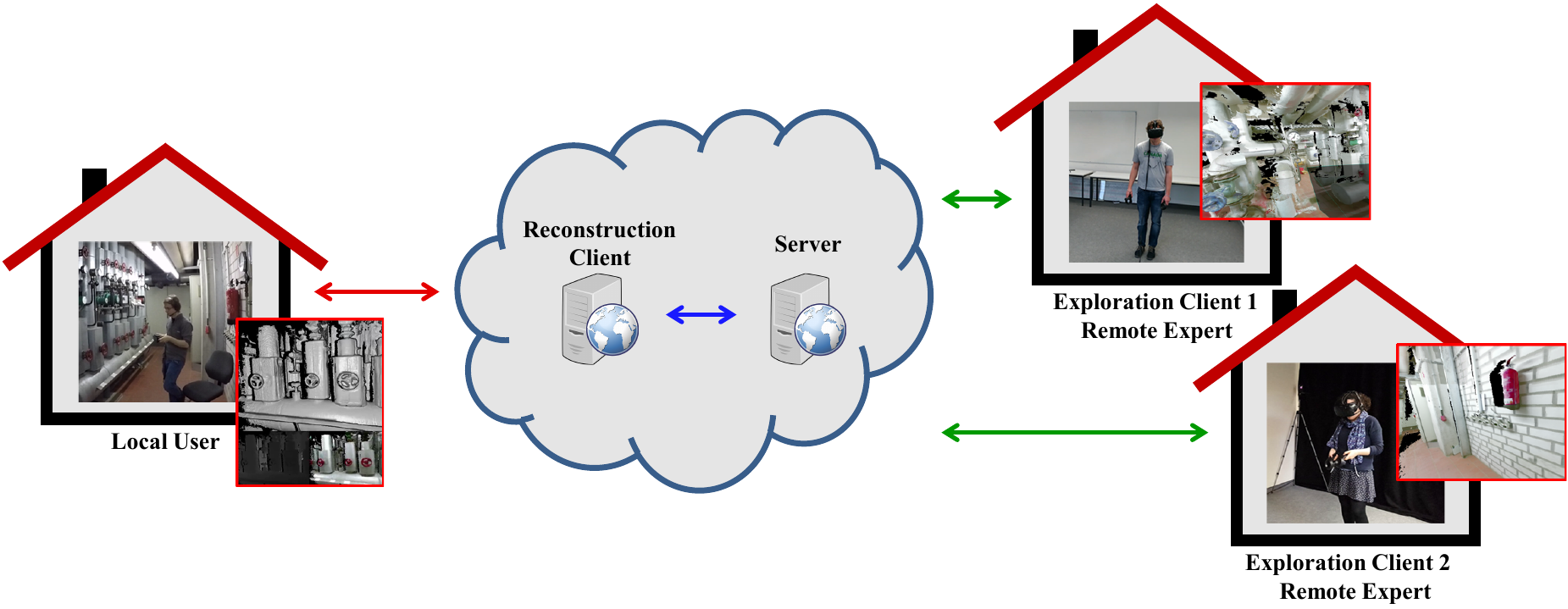}
    \caption{Illustration of our novel multi-client live telepresence framework for remote collaboration: RGB-D data captured with consumer-grade cameras represent the input to our real-time large-scale reconstruction technique that is based on a novel thread-safe GPU hash map data structure. Efficient data streaming is achieved by transmitting a novel compact representation of the reconstructed model in terms of Marching Cubes indices. Multi-client live telepresence is achieved by the server's independent handling of client requests.}
    \label{fig:teaser}
}

\vgtcinsertpkg

\begin{document}


\firstsection{Introduction}

\maketitle

%-------------------------------------------------------------------------
% Already defined in main template
%\section{Introduction}

One of the main motivations behind virtual reality research has always been to allow users to immersively and subjectively explore remote places or environments.
An experience of telepresence could benefit applications as diverse as remote collaboration, entertainment, advertisement, teaching, hazard site exploration, or rehabilitation. 
Thanks to advances in display technology and the emergence of high-resolution head-mounted devices, we have seen a recent surge in virtual reality solutions.
However, it has long been known that traditional display parameters like resolution, frame rate and contrast are not the only factors contributing to an immersive viewing experience.
The presentation of the data, its consistency, low-latency control to avoid motion sickness, the degree of awareness and the suitability of controller devices are just as important~\cite{Fontaine:1992,Held:1992,Witmer:1998}.
For applications such as remote exploration, remote collaboration or teleconferencing, these conditions are not easily met, as the scene is not pre-built but needs to be reconstructed on-the-fly from 3D input data captured by a person or robotic device.
At the same time, the data flow in a well-designed system should give multiple remote users the freedom to individually explore, for instance using head-mounted displays (HMD), the current state of reconstruction in the most responsive way possible.

A particular challenge, therefore, is to find a suitable coupling between the acquisition and viewing stages that respects the practical limitations imposed by available network bandwidth and client-side compute hardware while still guaranteeing an immersive exploration experience.
For this purpose, teleconferencing systems for transmitting dynamic 3D models of their \emph{users} typically rely on massive well-calibrated acquisition setups with several statically mounted cameras around the region of interest~\cite{Vasudevan:2011,Orts-Escolano:2016,Fairchild:2016}.
Instead, we direct our attention to the remote exploration of \emph{places} using portable, consumer-grade acquisition devices, for instance in scenarios of remote inspection or consulting.
On the acquisition site, a user digitizes their physical environment using consumer-grade 3D capture hardware.
Remote clients can perform immersive and interactive live inspection of that environment using off-the-shelf VR devices even while it is acquired and progressively refined.
In this scenario, additional challenges arise as the incoming amount of captured data may be high and may also significantly vary over time depending on the size of the scene that is currently imaged.
The latter particularly happens for strongly varying object distances within the captured data, whereas the amount of data over time remains in the same order of magnitude if the objects are within the same distance to the capturing camera (as met for teleconferencing scenarios).
A first attempt towards interactive virtual live inspection of real scenes~\cite{mossel} built upon real-time voxel block hashing based 3D reconstruction~\cite{infinitam} using implicit truncated signed distance fields (TSDFs) that has become a well-established method for high-quality reconstructions~\cite{kinectfusion,kinectfusion2,Whelan:2012,Newcombe:2015,Dou:2016,niessner,infinitam}.
Voxel blocks that are completely processed, \ie those that are no longer visible, are immediately sent to the remote client and locally converted into a mesh representation using Marching Cubes~\cite{Lorensen:1987:MCH} to perform the actual rendering.
Besides the fact that the system is restricted to one remote user, other limitations are the rather high bandwidth requirement of up to 175MBit/s and the missing handling of network failures where the remote client has to reconnect.
In particular for multi-client scenarios, handling both the bandwidth problem and the reconnection problem is of utmost importance to allow a satisfactory interaction between the involved users.

To overcome these problems, we propose a novel efficient low-cost multi-client remote collaboration system for the exploration of quasi-static scenes that is designed as a scalable client-server system which handles an arbitrary number of exploration clients under real-world network conditions (including the recovery from full outages) and using consumer-grade hardware.
The system consists of a voxel block hashing based reconstruction client, a server managing the reconstructed model and the streaming states of the connected clients as well as the exploration clients themselves (see \autoref{fig:teaser}).
The realization of the system relies on the following two key innovations:
\begin{itemize}[leftmargin=1em]\setlength\itemsep{0.0em}
    \item A novel scene representation and transmission protocol based on Marching Cubes (MC) indices enables the system to operate in low-bandwidth remote connection scenarios. Rather than reconstructing geometry on the server site or even performing server-side rendering, our system encodes the scene as a compressed sequence of voxel block indices and values, leaving the final geometry reconstruction to the exploration client. This results in significantly reduced bandwidth requirements compared to previous voxel based approaches~\cite{mossel}.
    \item For the scalable, reliable and efficient management of the streaming states of the individual exploration clients, we propose a novel thread-leveled GPU hash set and map datastructure that guarantees successful concurrent retrieval, insertion and removal of millions of entries on the fly while preserving key uniqueness without any prior knowledge about the data.
\end{itemize}

From a system point of view, the extension of the system towards multiple reconstruction clients~\cite{Golodetz2018Collaborative} is also envisioned but beyond the scope of this paper.
In order to overcome the inherently limited resolution of voxel-based scene representations, we also include a lightweight projective texture mapping approach that enables the visualization of texture details at the full resolution of the depth camera on demand.
Users collaboratively exploring the continuously captured scene experience a strong telepresence effect and are directly able to start conversation about the distant environment.
We motivate the need of a client server system, provide a discussion of the respective challenges and design choices, and evaluate the proposed system regarding latency, visual quality and accuracy.
Furthermore, we demonstrate its practicality in a multi-client remote servicing and inspection role-play scenario with non-expert users (see supplemental video).

%-------------------------------------------------------------------------
\section{Related Work}

\begin{figure*}[t]
    \centering
    \includegraphics[width=0.975\textwidth]{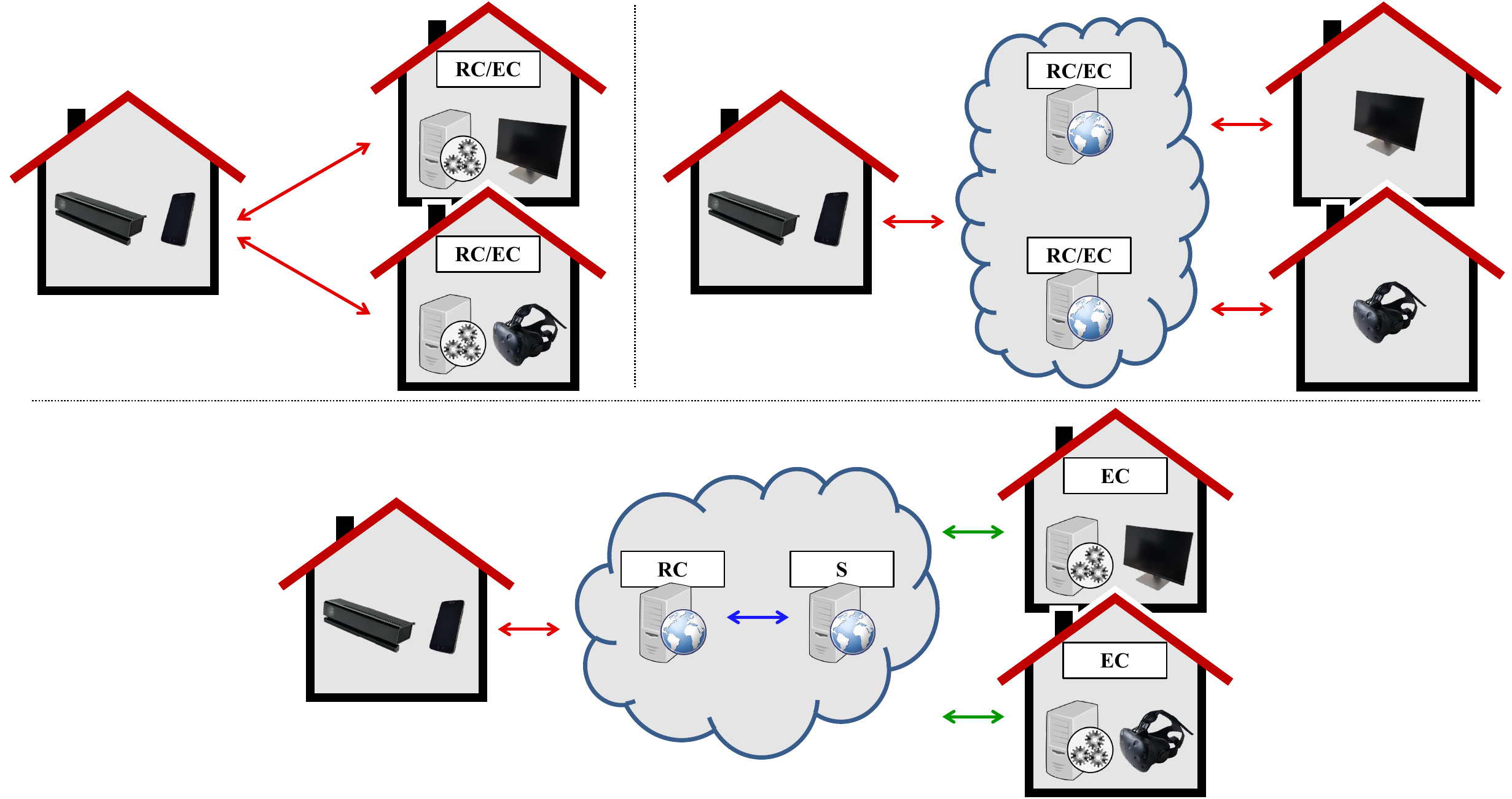}
    \caption{Possible design choices regarding the architecture of an end-to-end VR collaboration system. Although both the reconstruction client (RC) and the exploration client (EC) can be realized at the remote expert's side or inside the cloud (top row) to rely on standard video streaming techniques (red arrows), such systems either impose an extremely high computational burden to the client's machine (top left) or fail to provide an immersive VR experience due to Internet latencies. Our system (bottom) overcomes these limitations by streaming the reconstructed 3D model to the individual exploration clients using a novel compact and bandwidth-optimized representation (green arrows). Note, that the use of multiple reconstruction clients can be naturally realized in this setting by transmitting the original representation (blue arrows) between reconstruction client (RC) and server (S).}
    \label{fig:design_choices}
\end{figure*}

In this section, we provide an overview of previous efforts related to our novel large-scale, real-time 3D reconstruction and streaming framework for immersive multi-client telepresence categorized according to the developments regarding telepresence, 3D reconstruction and hashing.

\subsection{Telepresence}

Real-time 3D reconstruction is a central prerequisite for many immersive telepresence applications.
Early multi-camera telepresence systems did not allow the acquisition and transmission of high-quality 3D models in real-time to remote users due to limitations regarding the hardware at the time~\cite{Fuchs:1994,Kanade:1997,Mulligan:2000,Towles:2002,Tanikawa:2005,Kurillo:2008} or the applied techniques such as the lacking reconstruction accuracy of shape-from-silhouette approaches for concave surface regions~\cite{Petit:2010,Loop:2013}.
Then the spreading access to affordable commodity depth sensors such as the Microsoft Kinect led to the development of several 3D reconstruction approaches at room scale~\cite{kinectfusion,Maimone:2012,Maimone:2012b,Molyneaux:2012,Jones:2014,Fuchs:2014}.
However, the high sensor noise as well as temporal inconsistency in the reconstruction limited the quality of the reconstructions.
Furthermore, Photoportals~\cite{Kunert:2014} have been proposed to provide immersive access to pre-captured 3D virtual environments while also supporting remote collaborative exploration.
However, including live-captured contents comes at the cost of a significant lag as well as a reduced resolution.
In contrast, the Holoportation system~\cite{Orts-Escolano:2016} is built on top of the accurate real-time 3D reconstruction pipeline Fusion4D~\cite{Dou:2016} and involves real-time data transmission as well as AR and VR technology to achieve an end-to-end immersive teleconferencing experience.
However, massive hardware requirements, \ie several high-end GPUs running on multiple desktop computers, were needed to achieve real-time performance, where most of the expensive hardware components need to be located at the local user's side.
In the context of static scene telepresence, Mossel and Kr{\"o}ter \cite{mossel} developed an interactive single-exploration-client VR application based on current voxel block hashing techniques~\cite{infinitam}.
Although the system is restricted to only one exploration client, the bandwidth requirements of this approach have been reported to be up to 175MBit/s in a standard scenario.
A further issue resulting from the direct transmission of the captured data to the rendering client occurs in case of network interruptions where the exploration client has to reconnect to the reconstruction client.
Since the system does not keep track of the transmitted data, parts of the scene that are reconstructed during network outage will be lost.
While previous approaches are only designed for single client telepresence or do not support interactive collaboration, our approach overcomes these limitations and enables a variety of new applications.

\subsection{3D Reconstruction}

The key to success of the recently emerging high-quality real-time reconstruction frameworks is the underlying data representation that is used to fuse the incoming sensor measurements.
Especially the modeling of surfaces in terms of implicit truncated signed distance fields (TSDFs) has become well-established for high-quality reconstructions.
Earlier of these volumetric reconstruction frameworks such as KinectFusion~\cite{kinectfusion,kinectfusion2} rely on the use of a uniform grid so that the memory requirement linearly scales with the overall grid size and not with the significantly smaller subset of surface areas.
As this is impractical for handling large-scale scenes, follow-up work focused on the development of efficient data structures for real-time volumetric data fusion by exploiting sparsity in the TSDF.
This has been achieved based on using moving volume techniques~\cite{Roth:2012,Whelan:2015}, representing scenes in terms of blocks of volumes that follow dominant planes~\cite{Henry:2013} or height maps that are parameterized over planes~\cite{Schoeps:2014}, or using dense volumes only in the vicinity of the actual surface areas to store the TSDF~\cite{Chen:2013,niessner,infinitam}.
The allocated blocks that need to be indexed may be addressed based on tree structures or hash maps.
Tree structures model the spatial hierarchy at the cost of a complex parallelization and a time-consuming tree traversal which can be avoided with the use of hash functions that, however, discard the hierarchy.
Nie{\ss}ner \etal \cite{niessner} proposed real-time 3D reconstruction based on a spatial voxel block hashing framework that has been later optimized~\cite{infinitam}.
Drift that may lead to the accumulation of errors in the reconstructed model~\cite{niessner} can be counteracted by implementing loop closure~\cite{Kaehler:2016,Dai:2017}. 
Due to its efficiency, we built our remote collaboration system on top of the voxel block hashing approach and adapt the latter to the requirements discussed before.
Very recently, Golodetz \etal \cite{Golodetz2018Collaborative} presented a system for multi-client collaborative acquisition and reconstruction of static scenes with smartphones.
For each connected camera, a submap representing the client-specific scene part is reconstructed and managed by a server.
After capturing has finished, all submaps are merged into a final globally consistent 3D model to avoid artifacts arising from non-perfectly matching submap borders~\cite{Kaehler:2016}.
In contrast we focus on the development of a practical collaboration system for on-the-fly scene inspection and interaction by an arbitrary number of exploration clients.
In this scenario, issues such as the submap jittering caused by progressive relocalization during the capturing process have to be handled carefully in order to preserve an acceptable VR experience. As the respective adequate adjustment of the submaps has to be evaluated in the scope of comprehensive user studies, we consider this challenge to be beyond the scope of this paper.

\subsection{Hashing}

Lossless packing of sparse data into a dense map can be achieved via hashing.
However, developing such data structures on the GPU offering the reliability of their CPU-side counterparts is highly challenging.
Current voxel block hashing techniques~\cite{niessner,infinitam,Dai:2017} including hierarchical voxel block hashing~\cite{Kaehler:2016:Hierarchical} rely on the high camera frame rate to clean up block allocation failures in subsequent frames and, thus, guarantee consistent but not necessarily successful insertion and removal.
Only the guarantee regarding key uniqueness is strictly enforced to avoid that duplicate blocks are allocated and integrated during fusion.
Although data integration for some voxel blocks (and re-integration~\cite{Dai:2017}) might, hence, be staggered to a few subsequent frames, model consistency is still ensured by the high frame rate fusion.
To achieve a more reliable GPU hashing, perfect hashing approaches~\cite{Lefebvre:2006,Botelho:2013,Tran:2015} have been proposed that aim at collision-free hashing, but are hardly applicable for online reconstruction.
In the context of collision handling, minimizing the maximum age of the hash map, \ie the maximum number of required lookups during retrieval, by reordering key-value pairs similar to Cuckoo Hashing improves the robustness of the hash map construction~\cite{Garcia:2011}.
Similar to Alcantara \etal \cite{Alcantara:2011:Thesis}, who analyzed different collision resolving strategies, the entry size is restricted to 64-bit due to the limited support size of atomic exchange operations.
However, these approaches do not support entry removal and insertion is allowed to fail in case the defined upper bound on the maximum age is not achieved.
Stadium Hashing~\cite{Khorasani:2015} supports concurrent insertion and retrieval, but lacks removal, by avoiding entry reordering that would otherwise lead to synchronization issues.
Recently, Ashkinani \etal \cite{ashkiani2018dynamic} presented a fully dynamic hash map supporting concurrent insertion, retrieval, and also removal based chaining to resolve collisions.
However, their data structure cannot enforce key uniqueness, which is an essential property required by voxel block hashing frameworks to preserve model consistency.
In contrast, our hash map data structure overcomes all of the aforementioned limitations and is specifically suited for continuously updated reconstruction and telepresence scenarios.

%-------------------------------------------------------------------------
\section{Design Choices}
\label{sec:design_choices}

\begin{table*}[t]
    \footnotesize
    \centering
    \newcommand{\wC}{0.16\textwidth}
    \definecolor{bad}{RGB}{229,0,0}
    \definecolor{good}{RGB}{0,153,0}
    \caption{Advantages and disadvantages of different scene representations for remote collaboration systems.}
    \begin{tabular}{lcC{\wC}C{\wC}C{\wC}C{\wC}C{\wC}c}
        \toprule
        Data Representation  & Flexibility                  & Individual Exploration       & Re-Connection                & Data Management        & Compactness                 \\
        \midrule
        RGB-D Data           & \textcolor{bad}{-}           & \textcolor{bad}{-}           & \textcolor{bad}{-}           & \textcolor{good}{easy} & \textcolor{good}{good}      \\
        Voxel Block Model    & \textcolor{good}{\checkmark} & \textcolor{good}{\checkmark} & \textcolor{good}{\checkmark} & \textcolor{good}{easy} & \textcolor{bad}{bad}        \\
        Mesh                 & \textcolor{good}{\checkmark} & \textcolor{good}{\checkmark} & \textcolor{good}{\checkmark} & \textcolor{bad}{hard}  & \textcolor{good}{good}      \\
        MC index based Model & \textcolor{good}{\checkmark} & \textcolor{good}{\checkmark} & \textcolor{good}{\checkmark} & \textcolor{good}{easy} & \textcolor{good}{very good} \\
        \bottomrule
    \end{tabular}
    \label{tab:scene_representations}
\end{table*}

In a practical remote communication and collaboration system, users should be able to directly start a conversation about the -- possibly very large -- environment/scene and experience an immersive live experience without the need for time-consuming prerecording similar to a telephone call.
Such systems rely on efficient data representation and processing (see \autoref{tab:scene_representations}), immediate transmission as well as fast and compact data structures to allow reconstructing and providing a virtual 3D model in real time to remote users.
In order to meet the requirements regarding usability, latency, and stability, several crucial design choices have to be taken into account.
In particular, we thus focus on the discussion of a system design that benefits a variety of applications, while allowing the distribution of the computational burden according to the hardware availability respectively, \ie to the cloud or to the remote expert's equipment, and scaling to many remote clients.

\paragraph{Na{\"i}ve Input Video Streaming}

An obvious strategy for the interactive exploration of a live-captured scene by the user is the transmission of the RGB-D input sequence and the reconstruction of the scene model at the exploration client's site (see \autoref{fig:design_choices} top left).
Whereas the current state of the art in image and video compression techniques as well as real-time reconstruction would certainly be sufficient for the development of such systems, this approach has several limitations.
First, such a setup imposes an extremely high computational burden to the remote expert's machine, where both the reconstruction and the rendering have to be performed, such that a smooth VR experience at 90Hz may not be guaranteed.
Furthermore, in case of network outages, parts of the scene that are acquired while the exploration client is disconnected cannot be recovered automatically and the local user performing the capturing of the scene is forced to move back and acquire the missing parts again.
In the worst case where the exploration client completely looses the currently reconstructed model, \eg when the user accidentally closes the client, the whole capturing session must be restarted.
In contrast, this problem can be avoided by instead streaming parts of the fused 3D model where the streaming order is not limited to the acquisition order and can, thus, be controlled for each exploration client independently according to their particular interests.

\paragraph{Full Cloud Video Streaming}

Alternatively, the full reconstruction including triangulation could be performed on a central cloud server and only RGB-D video streams are transmitted from/to the users (see \autoref{fig:design_choices} top right).
While re-connections do not require further handling and data loss is no longer an issue, however, Internet latency becomes an apparent problem and prohibits an immersive VR experience.
Lags in transmitting the video data directly affect the user experience.
Standard approaches trying to compensate this issue rely on the view-adapted transmission of 360 degree video data (\eg \cite{Corbillon2017,Hosseini2016,Fan2017}).
This allows inspecting the scene based on head rotations, however, translations through the scene are not supported.
Furthermore, this not only requires that the users do not perform any fast movements, but also results in drastically increased bandwidth requirements due to the transmission of 360 degree video data which can easily result in the range of around 100MBit/s for SD video at 30Hz or more than 1GBit/s for 4K resolution at 120Hz respectively~\cite{Mangiante2017} which is higher than streaming the 3D model.
The additional use of techniques for view specification based on \eg fixation prediction~\cite{Fan2017} result in additional delays of around 40ms which represents a noticeable perceivable lag in remote collaboration scenarios and reduces the interactive experience.
In addition, when the reconstruction is finished or paused and the 3D model does not change for a certain time, the video stream of the renderings still requires a constantly high amount of bandwidth whereas the bandwidth required for streaming the 3D model would immediately drop to zero.

\paragraph{Mesh Data Streaming}

When deciding for the aforementioned server architecture, there remains still the question which data should be transferred from the server to the exploration clients.
Similar to full cloud-based video streaming, mesh updates could be streamed to the exploration clients and directly rendered at their machines using standard graphics APIs.
Whereas the mesh representation is more compact in comparison to the voxel block model that is used for reconstruction, the number of triangles in each updated block largely differs depending on the amount of surface inside resulting in significantly more complicated and less efficient data management, updating and transmission.
Furthermore, the vertex positions, which are given in the global coordinate system, are much harder to compress due to their irregular and arbitrary bit pattern.
Instead, we propose a novel bandwidth-optimized representation based on Marching Cubes indices (see \autoref{sec:server}) that is even more compact after compression due to its more regular nature.

\paragraph{Centralized Data Processing}

We focus on the development of a system that is particularly designed for collaboration tasks where users can explore and interact with the captured scene while at the same time being able to observe the other client's interactions.
For this purpose, a central server is placed between the individual clients to simplify the communication between clients and move shared computational work away from the clients.
Using a server avoids complicated and error-prone dense mesh networks between all the exploration clients.
Furthermore, it naturally facilitates the integration of multiple reconstruction clients and it allows lower hardware requirements at the exploration clients.
This, in turn, makes the system suitable for a much broader variety of users.
Powerful hardware, required for the scalability to a large number of clients, can be provided as practical cloud services or similar services (see \autoref{fig:design_choices} bottom).

\paragraph{Hash Data Structure}

Efficient data structures are crucial for efficiently and reliably managing the set of updated blocks for each connected exploration client as well as the scene model and therefore have to be adequately taken into account during the design phase.
For data management, fast and efficient retrieval of subsets as well as guaranteed modification through duplicate-free insertion and deletion, which both implicitly perform retrieval to ensure uniqueness, are strictly required to avoid data loss during transmission or redundant streaming of data.
In particular, the streaming states of each connected client, \ie the set of updated data that needs to be transmitted, must be maintained in real-time to avoid delays during live exploration.
Since the support for re-connections is a major feature of our telepresence system, these states will contain the list of blocks updated in the time while the connection was down or all blocks in case the client was closed accidentally by the user.
Selecting a subset (which involves retrieval and deletion) as well as filling the state (which should be duplicate-free to avoid redundant transmissions) should, hence, be performed as fast as possible in parallel on the GPU for which hash data structures are highly suitable and have been well-established (\eg \cite{niessner,infinitam}).
While recently developed hashing approaches work well with high-frame-rate online 3D reconstruction techniques, their lack of strong guarantees regarding hash operations make them hardly applicable to use cases with high reliability requirements such as telepresence systems.
Dispensing with the uniqueness guarantee would lead to redundantly transmitted data and, hence, wasted bandwidth whereas artifacts such as holes will occur when insertion, removal, and retrieval cannot be guaranteed and these blocks get lost during streaming from the reconstruction client until the exploration client.
With a novel hash map data structure that supports concurrent insertion, removal, and retrieval including key uniqueness preservation while running on a thread level, we directly address these requirements.
A detailed evaluation regarding run time and further relevant design choices are provided in the supplemental material.

%-------------------------------------------------------------------------
\section{Proposed Remote Collaboration System}
\label{sec:system_components}

\begin{figure*}[t]
    \centering
    \includegraphics[width=\textwidth]{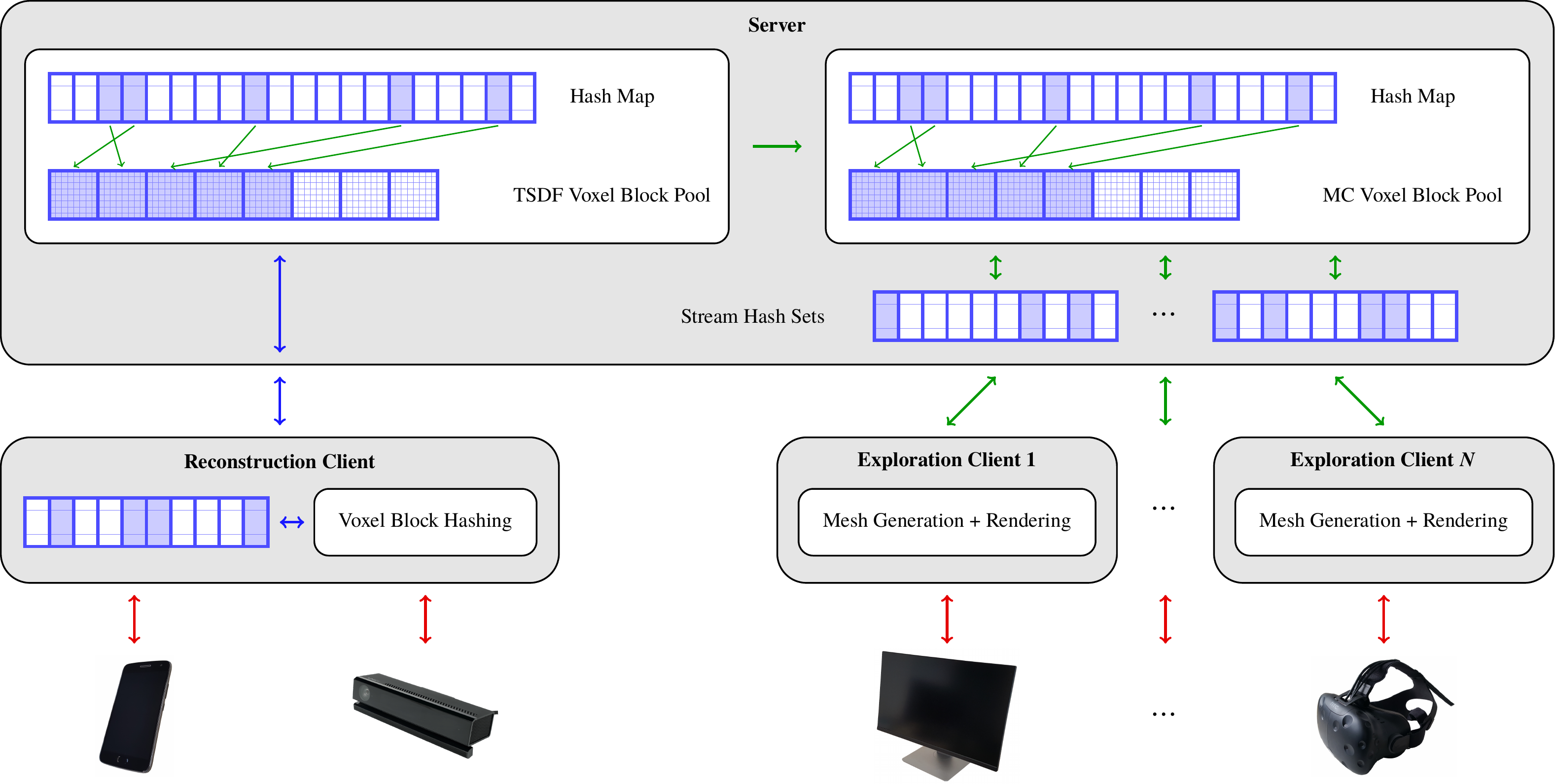}
    \caption{Our novel 3D reconstruction and streaming framework for multi-client remote collaboration. RGB-D images acquired by consumer cameras, \eg smartphones or the Kinect device, are streamed to the reconstruction client (red arrows) which updates the virtual model and transfers it to the server (blue arrows). The server converts the received data to a novel bandwidth-optimized representation based on Marching Cubes (MC) indices and manages a set of updated blocks that are queued for streaming for each connected exploration client. By design, our system supports an arbitrary number of exploration clients that can independently request the currently relevant updated parts of the model (green arrows) and integrate them into their locally generated mesh from which images are rendered in real-time and displayed on devices such as VR headsets or screens. For an immersive lag-free experience, the computational load during streaming is distributed using our novel hash map and set data structures. Red arrows are used to represent the image streaming, while blue and green arrows are used to represent the streaming of TSDF and MC voxel blocks.}
    \label{fig:streaming_pipeline}
\end{figure*}

\begin{figure*}[t]
    \centering
    \includegraphics[width=0.95\textwidth]{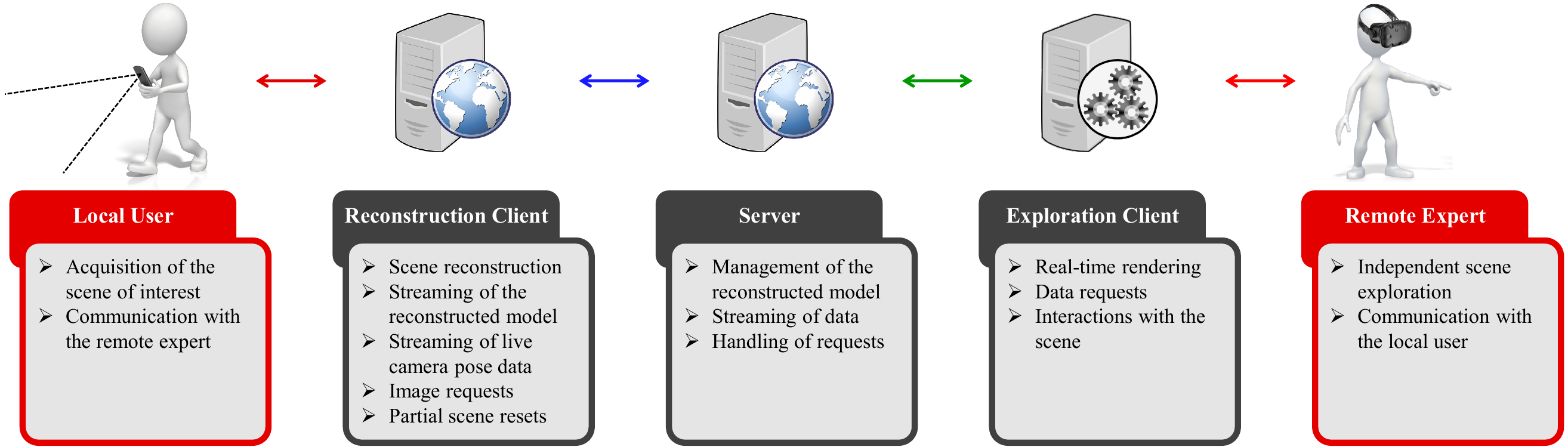}
    \caption{Components of our framework and their respective tasks. Images are partially provided by PresenterMedia \cite{PresenterMedia}.}
    \label{fig:PipelinewithTasks}
\end{figure*}

The overall server-client architecture of our novel framework for efficient large-scale 3D reconstruction and streaming for immersive remote collaboration based on consumer hardware is illustrated in \autoref{fig:streaming_pipeline} and the tasks of the involved components are shown in \autoref{fig:PipelinewithTasks}.
RGB-D data acquired with commodity 3D depth sensors as present in a growing number of smartphones or the Kinect device are sent to the reconstruction client, where the 3D model of the scene is updated in real time and transmitted to the server.
The server manages a copy of the reconstructed model, a corresponding, novel, bandwidth-optimized voxel block representation, and the further communication with connected exploration clients.
Finally, at the exploration client, the transmitted scene parts are triangulated to update the locally generated mesh which can be immersively explored \ie with VR devices.
Clients can connect at any time before or after the capturing process has started.
In the following sections, we provide more detailed descriptions of the individual components of our framework, \ie the reconstruction client, the server, and the exploration client, which is followed by an in-depth discussion of the novel data structure (see \autoref{sec:data_structure}).
Additional implementation details for each component are provided in the supplemental material.

\subsection{Reconstruction Client}
\label{sec:reconstruction_client}

The reconstruction client receives a stream of RGB-D images acquired by a user and is responsible for the reconstruction and streaming of the virtual model.
We use voxel block hashing~\cite{niessner,infinitam} to reconstruct a virtual 3D model from the image data.
Since the bandwidth is limited, the as-efficient-as-possible data handling during reconstruction is of great importance.
For this purpose, we consider only voxel blocks that have already been fully reconstructed and for which no further immediate updates have to be considered, \ie blocks that are not visible in the current sensor's view anymore and have been streamed out to CPU memory~\cite{mossel}.
In contrast, transmitting blocks that are being still actively reconstructed and, thus, will change over time which results in an undesirable visualization experience for exploration clients.
Furthermore, continuously transmitting these individual blocks during the reconstruction process results in extremely increasing bandwidth requirements which make this approach infeasible to real-world scenarios.
In contrast to Mossel and Kr{\"o}ter \cite{mossel}, we concurrently insert the streamed-out voxel blocks into a hash set which allows us to control the amount of blocks per package that are streamed and avoids lags by distributing the work across multiple frames similar to the transfer buffer approach of the InfiniTAM system~\cite{infinitam}.
To mitigate the delay caused by transmitting only fully reconstructed parts of the scene, we add the currently visible blocks at the very end of the acquisition process as well as when the user stops moving during capturing or the hardware including the network connection are powerful enough to stream the complete amount of queued entries.
In particular, we check whether the exponential moving average (EMA) of the stream set size over a period of $ \tau $ = 5 seconds~\cite{zumbach2001operators} is below a given threshold and the last such prefetching operation is at least 5 seconds ago.
The EMA is updated as
\begin{equation}
    EMA_\tau^{(t_{n + 1})} = u \, EMA_\tau^{(t_{n})} + (v - u) \, s_{n} + (1 - u) \, s_{n + 1}
\end{equation}
with
\begin{equation}
    u = e^{-a}
    , \quad
    v = \frac{1 - u}{a}
    , \quad
    a = \frac{t_{n + 1} - t_{n}}{\tau}.
\end{equation}
This ensures that the delayed but complete model is available to the server and the exploration clients at all times.
After fetching a subset of stream set (via concurrent removal) and the respective voxel data from the model, we compress them using lossless compression~\cite{Collet:2017:zstd} and send them to the server.
In addition to the pure voxel data, the reconstruction client and the exploration clients send their camera intrinsics and current camera pose to the server where they are forwarded to each connected exploration client to enable interactive collaboration.
Furthermore, requests for high-resolution textures on the model by the exploration clients, required \eg for reading text or measurement instruments, are handled by transmitting the sensor's current RGB image to the reconstruction client where it is forwarded to the server and the exploration clients.
To make our framework also capable of handling quasi-static scenes, where the scene is allowed to change between two discrete timestamps, as \eg occurring when an instrument cabinet has to be opened before being able to read the instruments, our framework also comprises a reset function that allows the exploration client to request scene updates for selected regions.
This can be achieved by deleting the reconstructed parts of the virtual model that are currently visible and propagating the list of these blocks to the server.

\subsection{Server}
\label{sec:server}

The server component is responsible for managing the global voxel block model and the list of queued blocks for each connected exploration client.
Furthermore, it converts incoming TSDF voxel blocks into our novel MC voxel block representation.
Finally, it forwards messages between clients and distributes camera and client pose data for an improved immersion and client interaction.

In order to reduce the computational burden and infrastructural requirements regarding network bandwidth, the streamed data should be as compact as possible while being efficiently to process.
Instead of streaming the model in the original TSDF voxel block representation of the voxel block hashing technique~\cite{mossel} to the exploration clients, we compute and transmit a bandwidth-optimized representation based on Marching Cubes~\cite{Lorensen:1987:MCH}.
Thus, a TSDF voxel (12~bytes), composed of a truncated signed distance field (TSDF) value (4~bytes), a fusion weight (4~bytes), and a color (3~bytes +~1~byte alignment), is reduced to a MC voxel, \ie a Marching Cubes index (1~byte), and a color value (3~bytes).
Furthermore, we cut off those voxel indices $ i $ and colors $ c $ where no triangles will be created, \ie for
\begin{equation}
    \mathcal{S}^{c} = \left\{ (i, c) \mid i = 0 \lor i = 255 \right\},
\end{equation}
by setting the values $ i $ and $ c $ to zero.
While omitting the interpolation weights, resulting in lossy compression, might seem drastic in terms of reconstruction quality, we show that the achieved improvement regarding compression ratio and network bandwidth requirement outweigh the slight loss of accuracy in the reconstruction (see \autoref{sec:evaluation}).
Compared to a binary representation of the geometry that would lead to the same quality and a similar compression ratio, our MC index structure directly encodes the triangle data and enables the independent and parallel processing at the remote site by removing neighborhood dependencies.

Incoming data sent by the reconstruction client are first concurrently integrated into the TSDF voxel block model and then used to update the corresponding blocks and their seven neighbors in negative direction in the MC voxel block representation.
Updating the neighbors is crucial to avoid cuts in the mesh due to outdated and inconsistent MC indices.
To avoid branch divergence and inefficient handling of special cases, we recompute the whole blocks instead of solely recomputing the changed parts.
The list of updated MC voxel blocks is then concurrently inserted to each exploration client's stream hash set.
Maintaining such a set for each connected client not only enables advanced streaming strategies required for a lag-free viewing experience (see \autoref{sec:exploration_client}).
It also allows them to reconnect at any point in time, \eg after network outages, and still explore the entire model since their stream sets are initially filled with the complete list of voxel blocks via concurrent insertion.
After selecting all relevant blocks, a random subset of at most the request size limit is extracted via concurrent removal and the corresponding voxel data are retrieved, compressed~\cite{Collet:2017:zstd} and sent to the exploration client.

\subsection{Exploration Client}
\label{sec:exploration_client}

The exploration client's tasks comprise generating surface geometry from the transmitted compact representation in terms of MC indices, updating the current version of the reconstructed model at the remote site, and the respective rendering of the model in real-time.
Therefore, exploration clients are allowed to request reconstructed voxel blocks according to the order of their generation during reconstruction, depending on whether they are visible in the current view of the client, or in a random order which is particularly useful in the case when the currently visible parts of the model are already complete, and thus, other parts of the scene can be prefetched.
Since the exploration client controls the request rate and size, a lag-free viewing experience is achieved by adapting these parameters depending on the client's hardware resources.

The received MC voxel blocks are decompressed in a dedicated thread, and the block data is passed to a set of reconstruction threads which generate the scene geometry from the MC indices and colors of the voxels.
We reduce the number of draw calls to the graphics API by merging $ 15^3 $ voxel blocks into a mesh block instead of rendering each voxel block separately~\cite{mossel}. 
To reduce the number of primitives rendered each frame, we compute three level of details (LoDs) from the triangle mesh, where one voxel, eight voxels or 64 voxels respectively are represented by a point and the point colors are averaged over the voxels.
During the rendering pass, all visible mesh blocks are rendered, while their LoD is chosen according to the distance from the camera.
We refer to the supplemental material for more details.

To allow a better interaction between the involved clients, each exploration client additionally sends its own pose to the server, which distributes it to other exploration clients, so that each user can observe the poses and movements of other exploration clients within the scene. 
Analogously, the current pose of the reconstruction client is visualized in terms of the respectively positioned and oriented camera frustum.
Furthermore, users can interactively explore the reconstructed environment beyond pure navigation by measuring 3D distances between interactively selected scene points.
For the purpose of depicting structures below the resolution of the voxel hashing pipeline as \eg required for reading measurement instruments or texts, the exploration client can send requests to the server upon which the RGB image currently captured by the sensor is directly projected onto the respective scene part and additionally visualized on a virtual measurement display.

%-------------------------------------------------------------------------
\section{Hash Map and Set Data Structures}
\label{sec:data_structure}

\begin{figure}[t]
    \centering
    \includegraphics[width=\columnwidth]{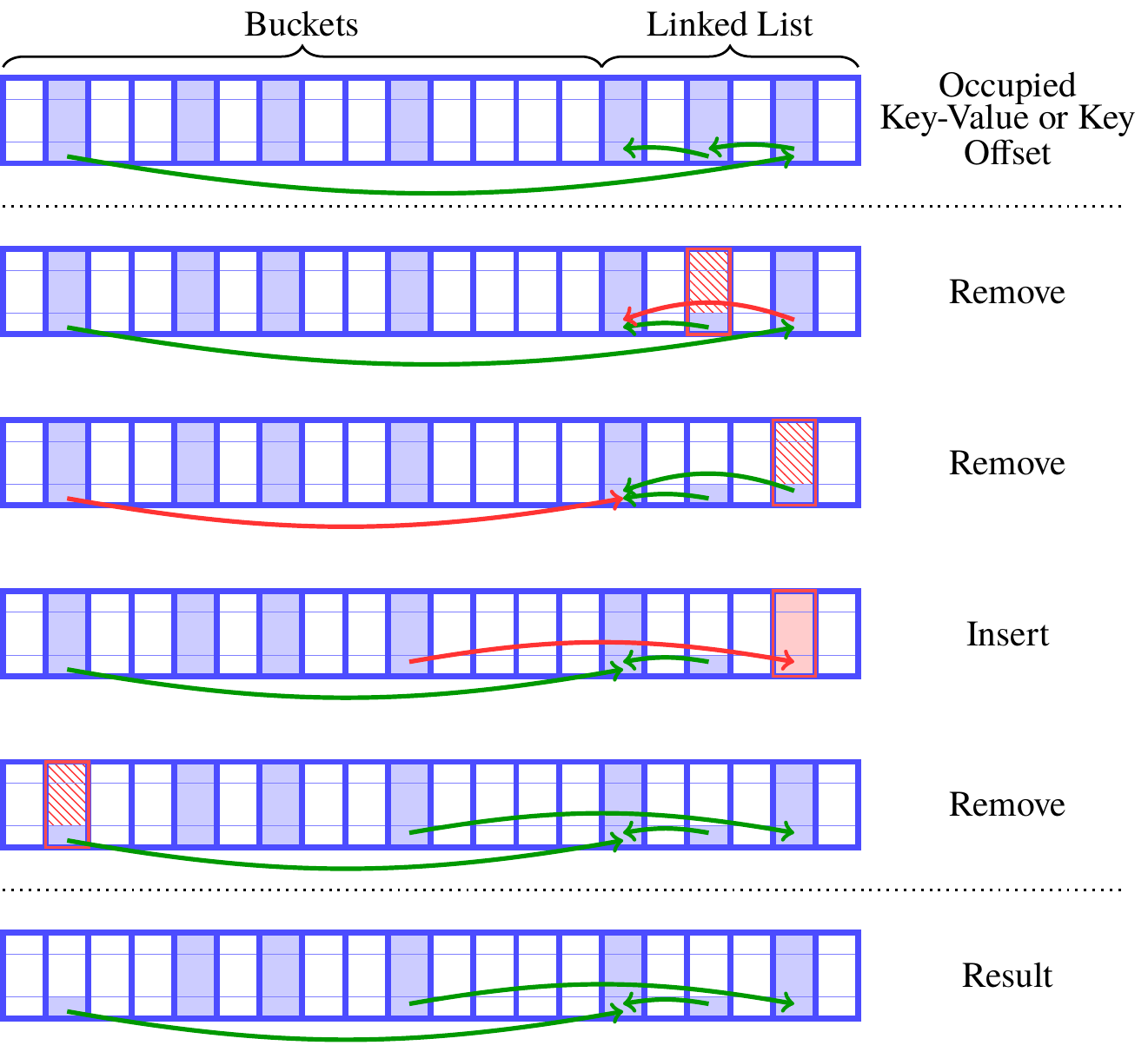}
    \caption{Illustration of thread-safe hash map/set modifications on the GPU by maintaining the proposed invariant. The importance of thread safety has its origin in the guarantees for successful concurrent retrieval, insertion and removal while preserving key uniqueness. This figure depicts one possible order for the operations to resolve the requested task when processing four operations in parallel. In the resulting structure, dead links and empty buckets might occur which, however, are not problematic and automatically cleaned up during further operations.}
    \label{fig:hash_map_set_insert_erase}
\end{figure}

For the purpose of large-scale 3D reconstruction and streaming to an arbitrary number of remote exploration clients, we developed a thread-safe GPU hash data structure allowing fast and simple management including dynamic concurrent insertion, removal and retrieval of millions of entries with strong success guarantees.
In comparison to pure 3D reconstruction, maintaining consistency in multi-client telepresence is much more challenging since streaming data between clients requires that updates are not lost \eg due to synchronization failures.
Whereas previous approaches either allow failures~\cite{niessner,infinitam,Garcia:2011} or do not ensure key uniqueness~\cite{Khorasani:2015,ashkiani2018dynamic}, our robust hash data structure is not limited in this regard and represents the key to realize our real-time remote collaboration system.
A detailed evaluation in terms of design choices and runtime performance can be found in the supplemental material.

\paragraph{General Design}

Our streaming pipeline is built upon two different hash data structures.
The server and the individual client components use an internal map structure, that stores unique keys and maps a value to each of them, whereas the server-client streaming protocol relies on a set structure, which only considers the keys.
Thus, the major difference lies in the kind of stored data whereas the proposed algorithm for retrieval, insertion and removal is shared among them.
We built upon the single-entry data structure by K{\"a}hler \etal \cite{infinitam} which stores the values, \ie key-value pairs for the map structure (voxel block hashing and server model) and keys for the set (streaming states, see \autoref{fig:streaming_pipeline}) into a linear array.
Collisions are resolved through linked lists using per-entry offsets to the next elements and a stack structure that maintains the set of available linked list entries.
Voxel block hashing based reconstruction approaches rely on the high camera frame rate to clean up block allocation failures in subsequent frames~\cite{niessner,infinitam,Dai:2017,Kaehler:2016:Hierarchical} and, therefore, reduce synchronization to a minimum.
In contrast, failures in our telepresence system result in data loss during data transmission which cannot be recovered.
Thus, we need additional indicators to determine whether an entry is occupied and locks for synchronization to handle cases where several threads attempt to modify the same entry simultaneously.
Furthermore, we maintain a strong invariant which is required to achieve correct concurrency on the thread-level: \emph{At any time, the entry positions and the links to colliding values are preserved.}
\autoref{fig:hash_map_set_insert_erase} demonstrates mixed insertion and removal operations on our thread-safe hash data structure.
Detailed descriptions and implementation details of the hash and stack data structures as well as further design remarks are provided in the supplemental material.

\paragraph{Retrieval}

Since our proposed invariant ensures that entry positions are not allowed to change, finding an element in the hash map or set can be safely implemented as a read-only operation.
First, the bucket $ b $ of a given key value is computed according to the underlying hashing function.
In case of spatial hashing, this function could be defined as
\begin{equation}
    b = \left( x \cdot p_1 \oplus y \cdot p_2 \oplus z \cdot p_3 \right) \bmod n
\end{equation}
where $ (x, y, z) $ are the voxel block coordinates, $ p_1 = 73856093, p_2 = 19349669, p_3 = 83492791 $ represent prime numbers, and $ n $ denotes the number of buckets~\cite{niessner,infinitam}.
We check whether the entry is occupied and its key matches the query.
If both conditions are met, we found the key and return the current position.
Otherwise, we traverse the linked list through the offsets and check each entry in a similar manner.

\paragraph{Insertion}

For successful concurrent insertion, the modification of an entry by several threads needs to be handled while avoiding deadlocks.
We handle the latter problem by by looping over a non-blocking insertion function, which is allowed to fail, until the value is found in the data structure.
In the non-blocking version, we first check if the value is already inserted (by performing retrieval).
If the entry is not found, there are two possible scenarios: The value can be inserted at the bucket (if this entry is not occupied) or at the end of the bucket's linked list.
In both cases, other threads might attempt to also modify the entry at the same time.
This not only requires locking (which might fail to prevent deadlocks), but also a second occupancy check.
If both the lock is successfully acquired and the entry is still free, the value is stored and the entry is marked as occupied and unlocked.
In case the bucket was initially occupied (second scenario), we first find the end of the linked list by traversing the offsets and lock that entry.
Afterwards, we extract a new linked list position from the stack, store the value there, set the occupancy flag and reset its offset to zero.
Note that the offset is intentionally not reset in the removal operation to avoid a race condition (see the section below for details).
Finally, the offset to the new linked list entry is stored and the acquired lock is released.

\paragraph{Removal}

Removing elements as required when selecting voxel blocks for client-server streaming, is similar to insertion and also involves double checking during lock acquisition as well as looping over a non-blocking version.
Again, there are two possible scenarios: The entry may be located at the bucket or inside the linked list.
In the former case, we try to acquire the lock and then reset the value and mark the entry as unoccupied.
In contrast to the approach by Nie{\ss}ner \etal \cite{niessner}, the first linked list entry is not moved to the bucket to preserve our invariant.
Threads that try to erase this value might, otherwise, fail to find it.
We evaluated the impact of this change and observed that runtime performance was not affected.
If the value is inside the linked list (second scenario), we first find the previous entry and lock both entries.
Afterwards, the current entry is reset and marked as unoccupied, the offset of the previous entry is updated, and both locks are finally released.
As mentioned earlier, the offset is kept to avoid a race condition where other threads concurrently performing direct or indirect retrieval (inside insertion and removal) might not be able to access the remainder of the linked list which would lead to failures in all three operations.
Thus, we avoid the need for additional synchronization in the retrieval operation by delaying this step to the insertion operation.

%-------------------------------------------------------------------------
\section{Evaluation}
\label{sec:evaluation}

\begin{table*}[t]
    \footnotesize
    \centering
    \newcommand{\wC}{0.0875\textwidth}
    \caption{Bandwidth measurements of our system for various scenes. We compared mean (and maximum) bandwidths of our optimized MC voxel structure with 128-1024 blocks/request and 100Hz request rate to the standard TSDF representation with 512 blocks/request and unlimited rate. Across all scenes, our optimized representation saved more than 90\% of the bandwidth and scales linearly with the package size.}
    \begin{tabular}{lcC{\wC}C{\wC}C{\wC}C{\wC}C{\wC}c}
        \toprule
        Dataset   & Voxel Size [mm] &     \multicolumn{5}{c}{Bandwidth [MBit/s]}    & Model Size [\# Voxel Blocks] \\
                   &                 & MC 128      & MC 256      & MC 512      & MC 1024     & TSDF 512 & \\
        \midrule                     
        \emph{heating\_room} & 5                   &  4.5  (8.0) &  8.8 (12.3) & 17.5 (30.9) & 32.7 (71.3) & 561.5 (938.8) & 897 $ \times 10^3 $ \\
        \emph{pool}          & 5                   &  4.6  (7.1) &  9.0 (14.0) & 17.8 (29.7) & 29.3 (54.5) & 489.3 (937.0) & 637 $ \times 10^3 $ \\
        \emph{fr1/desk2}     & 5                   &  8.1 (11.6) & 16.2 (23.8) & 32.6 (46.8) & 61.0 (95.0) & 764.0 (938.6) & 134 $ \times 10^3 $ \\
        \emph{fr1/room}      & 5                   & 12.3 (23.6) & 16.4 (23.6) & 32.1 (42.2) & 57.6 (87.9) & 739.7 (938.0) & 467 $ \times 10^3 $ \\
        \emph{heating\_room} & 10                  &  5.1  (7.6) &  9.2 (14.4) & 14.6 (27.8) & 20.2 (63.7) & 216.8 (937.1) & 147 $ \times 10^3 $ \\
        \emph{pool}          & 10                  &  5.6  (8.5) &  9.9 (16.0) & 13.6 (27.2) & 16.9 (52.3) & 176.3 (937.0) & 104 $ \times 10^3 $ \\
        \emph{fr1/desk2}     & 10                  &  8.7 (11.2) & 14.3 (21.8) & 19.6 (39.2) & 24.4 (71.3) & 170.1 (436.4) &  23 $ \times 10^3 $ \\
        \emph{fr1/room}      & 10                  &  9.2 (12.5) & 15.7 (23.5) & 22.9 (46.1) & 28.5 (88.8) & 207.8 (936.6) &  86 $ \times 10^3 $ \\
        \bottomrule
    \end{tabular}
    \label{tab:bandwidth}
\end{table*}

\begin{table*}[t]
    \footnotesize
    \centering
    \newcommand{\wC}{0.0875\textwidth}
    \caption{Time measurements of our system for various scenes. We compared the time to stream the whole model represented by our optimized MC voxel structure with 128-1024 blocks/request and 100Hz request rate to the standard TSDF representation with 512 blocks/request and unlimited rate. The reconstruction speed is given by TSDF 512 and serves as a lower bound. For a voxel resolution of 5mm, a package size of 512 voxel blocks results in the best trade-off between required bandwidth and total streaming time. Increasing the size leads to slightly better results with less latency, but substantially higher bandwidths. For a resolution of 10mm, the optimal streaming time is reached with even smaller package sizes.}
    \begin{tabular}{lcC{\wC}C{\wC}C{\wC}C{\wC}C{\wC}c}
        \toprule
        Dataset   & Voxel Size [mm] & \multicolumn{5}{c}{Time [min]} & Model Size [\# Voxel Blocks] \\
                   &                 &  MC 128 &  MC 256 &  MC 512 & MC 1024 &  TSDF 512 & \\
        \midrule                     
        \emph{heating\_room} & 5               &    4:06 &    3:08 &    2:40 &    2:32 &      2:31 & 897 $ \times 10^3 $ \\
        \emph{pool}          & 5               &    2:14 &    1:32 &    1:12 &    1:09 &      1:08 & 637 $ \times 10^3 $ \\
        \emph{fr1/desk2}     & 5               &    0:39 &    0:31 &    0:27 &    0:24 &      0:22 & 134 $ \times 10^3 $ \\
        \emph{fr1/room}      & 5               &    1:46 &    1:14 &    1:01 &    0:57 &      0:56 & 467 $ \times 10^3 $ \\
        \emph{heating\_room} & 10              &    1:49 &    1:44 &    1:44 &    1:44 &      1:44 & 147 $ \times 10^3 $ \\
        \emph{pool}          & 10              &    0:54 &    0:50 &    0:50 &    0:50 &      0:50 & 104 $ \times 10^3 $ \\
        \emph{fr1/desk2}     & 10              &    0:21 &    0:19 &    0:19 &    0:19 &      0:18 &  23 $ \times 10^3 $ \\
        \emph{fr1/room}      & 10              &    0:46 &    0:42 &    0:41 &    0:41 &      0:41 &  86 $ \times 10^3 $ \\
        \bottomrule
    \end{tabular}
    \label{tab:timings}
\end{table*}

After providing implementation details, we perform an analysis regarding bandwidth requirements and the visual quality of our compact scene representation.
This is accompanied by the description of the usage of our framework in a live remote collaboration scenario as well as a discussion of the respective limitations.

\subsection{Implementation}

We implemented our framework using up to four desktop computers taking the roles of one reconstruction client, one server, and two exploration clients.
Each of the computers has been equipped with an Intel Core i7-4930K CPU and 32GB RAM.
Furthermore, three of them have been equipped with a NVIDIA GTX 1080 GPU with 8GB VRAM, whereas the fourth computer made use of a NVIDIA GTX TITAN X GPU with 12GB VRAM.
For acquisition, we tested two different RGB-D sensors by using the Microsoft Kinect v2, which delivered data with a resolution of 512 $ \times $ 424 pixels at 30Hz, and by using an ASUS Zenfone AR, which captured RGB-D data with a resolution of 224 $ \times $ 172 pixels at 10Hz.
Although the ASUS device is, in principle, capable of performing measurements at frame rates of 5-15Hz, we used 10Hz as a compromise between data completeness and speed.
Each of the exploration client users was equipped with an HTC Vive HMD with a native resolution of 1080 $ \times $ 1200 pixels per eye whereas the recommended rendering resolution (reported by the VR driver) is 1512 $ \times $ 1680 pixels per eye, leading to a total resolution of 3024 $ \times $ 1680 pixels.
Please note that the higher recommended resolution (in comparison to the display resolution) originates from the lens distortion applied by the VR system.
All computers were connected via a local network.

\subsection{Bandwidth and Latency Analysis}

\begin{figure}[t]
    \centering
    \includegraphics[width = \columnwidth, height = 0.5\textheight, keepaspectratio]{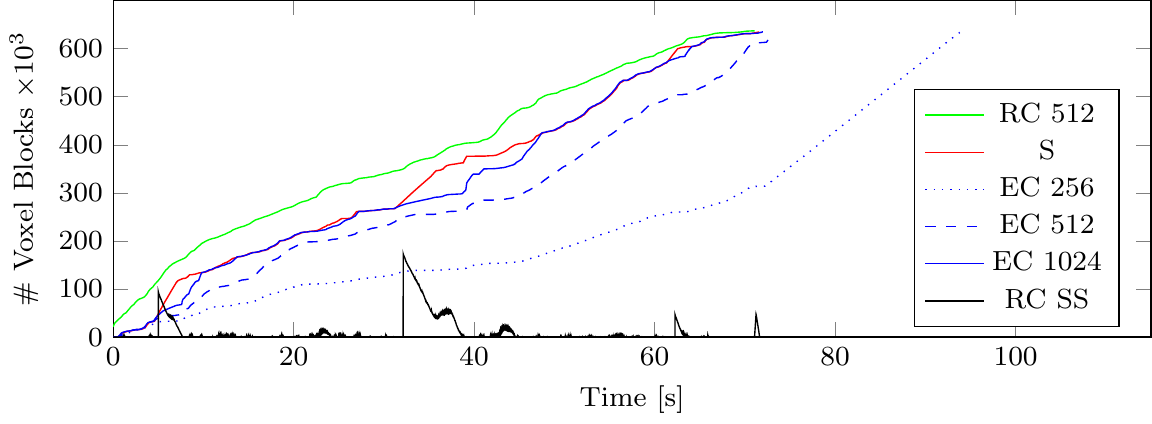}
    \caption{Streaming progress over time for the \emph{pool} dataset. Larger package sizes reduce the total transmission time of the virtual model to the exploration client (EC). To save bandwidth, only fully reconstructed blocks are streamed from the reconstruction client (RC) to the server (S) causing a noticeable delay, which becomes smaller when our prefetching queues the currently visible scene parts to the RC's stream set (RC SS).}
    \label{fig:completeness_over_time}
\end{figure}

\begin{figure*}[t]
    \centering
    \subfigure[Bandwidth requirements between reconstruction client (RC) and server (S) with 512 blocks/request and an exploration client (EC) with 256 blocks/request. As both RC and S are within the same network (i.e. in the cloud) in the proposed system architecture, the shown bandwidth requirements are still acceptable.]
    {
        \centering
        \includegraphics[width = 0.48\textwidth, height = 0.5\textheight, keepaspectratio]{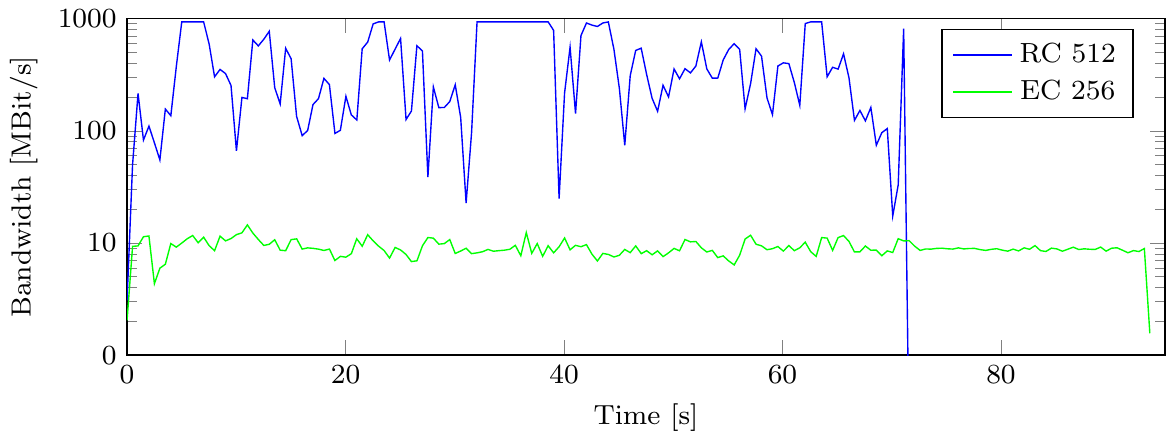}
        %\label{fig:bandwidth_over_time_rc_mc}
    }
    \hfill
    \subfigure[Bandwidth requirements between server (S) and exploration client (EC) with package sizes of 256, 512, and 1024 blocks/request.]
    {
        \centering
        \includegraphics[width = 0.48\textwidth, height = 0.5\textheight, keepaspectratio]{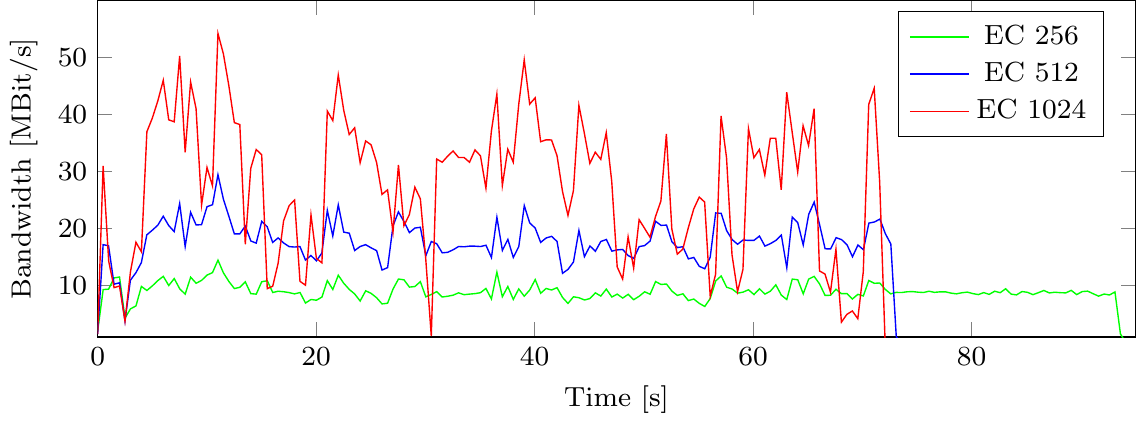}
        %\label{fig:bandwidth_over_time_mc}
    }
    \caption{Bandwidth measurements of our system over time for the \emph{pool} dataset.}
    \label{fig:bandwidth_over_time}
\end{figure*}

In the following, we provide a detailed quantitative evaluation of the bandwidth requirements of our novel collaboration system.
For the purpose of comparison, we recorded two datasets \emph{heating\_room} and \emph{pool} (see supplemental material) with the Kinect v2, and also used two further publicly available standard datasets that were captured with the Kinect v1~\cite{sturm12iros}.
Throughout the experiment, we loaded a dataset and performed the reconstruction on the computer equipped with the NVIDIA GTX TITAN X.
The model is then streamed to the server (second computer) and further to a benchmark client (third computer).
Compared to the exploration client, the benchmark client is started simultaneously to the reconstruction client, requests voxel blocks with a fixed predefined frame rate of 100Hz, and directly discards the received data to avoid overheads.
Using this setup, we measured the mean and maximum bandwidth required for streaming the TSDF voxel block model from the reconstruction client to the server and the MC voxel block model from the server to the benchmark client.
Furthermore, we also measured the time until the model has been completely streamed to the benchmark client.
For the voxel block hashing pipeline, we used 5mm and 10mm for the voxel size, 60mm for the truncation region and hash maps with $ 2^{20} $ and $ 2^{22} $ buckets as well as excess list sizes matching the respective active GPU and passive CPU voxel block pool sizes of $ 2^{19} $ and $ 2^{20} $ blocks.
The server and reconstruction client used the passive parameter set for their hash maps and sets.
The results of our experiment are shown in \autoref{tab:bandwidth} and \autoref{tab:timings}.
A further evaluation regarding the server scalability is provided in the supplemental material.

Across all scenes and voxel sizes, the measured mean and maximum bandwidths for our novel MC voxel structure scale linearly with the package size and are over one order of magnitude smaller compared to the standard TSDF voxel representation.
We measured higher bandwidths at 10mm voxel size than at 5mm for package sizes of 128 and 256 blocks.
Our stream hash set automatically avoids duplicates, which saves bandwidth in case the system works at its limits and can be considered as an adaptive streaming.
At 10mm this triggers substantially less and thus, more updates are sent to the server and exploration clients.
We also observed by a factor of two larger bandwidths for the datasets captured with the Kinect v1 in comparison to the ones recorded by us with the Kinect v2.
This is mainly caused by the lower reliability of the RGB-D data which contains more sensor noise as well as holes, which, in turn, results in a larger number of allocated voxel blocks that need to be streamed.
Furthermore, the faster motion induces an increased motion blur within the images, and thus leads to larger misalignments in the reconstructed model as well as even more block allocations.
However, this problem is solely related to the reconstruction pipeline and does not affect the scalability of our collaboration system.

The overall system latency is determined by the duration until newly seen parts of the scene are queued for transmission, \ie until they are streamed out to CPU memory, the latency of the network, and the package size of the exploration client's requests.
Since the whole system runs in real-time, \ie data are processed in the order of tens of milliseconds, the runtime latency within the individual components has a negligible impact on the total latency of the system.
In order to evaluate the bandwidth requirements and the overall latency, we performed further measurements as depicted in \autoref{fig:bandwidth_over_time} and \autoref{fig:completeness_over_time}.
Whereas the bandwidth for transmitting the TSDF voxel block representation has a high variance and ranges up to our network's limit of 1Gbit/s, our bandwidth optimized representation has not only lower requirements, \ie a reduction by more than 90\%, but also a significantly lower variance.
For a package size of 256 blocks, the model is only slowly streamed to the exploration client which results in a significant delay until the complete model has been transmitted.
Larger sizes such as 512 blocks affect both the mean bandwidth and the variance while further increases primarily affect the variance since less blocks than the package size need to be streamed (see \autoref{fig:bandwidth_over_time}).
This effect also becomes apparent in \autoref{fig:completeness_over_time} where lower package sizes lead to a smooth streaming and larger delays whereas higher values reduce the latency.
Furthermore, the delay between the reconstruction client and the server in the order of seconds is directly related to our choice of only transmitting blocks that have been streamed out to save bandwidth.
Note that directly streaming the actively reconstructed voxel blocks is infeasible due to extremely increasing bandwidth requirements (see Section~\ref{sec:reconstruction_client}).
Once our automatic streaming of the visible parts triggers, which can be seen in the rapid increases of the RC's stream set (RC SS), the gap between the current model at the reconstruction client and the streamed copy at the server becomes smaller.
Since the visible blocks are streamed in an arbitrary order, this results in lots of updates for already existing neighboring MC voxel blocks at the server site that need to be streamed to the exploration client.
Therefore, the exploration client's model grows slower than the server's model but this gap is closed shortly after the server received all visible blocks.
Note that the effects of this prefetching approach can be also seen in the reconstruction client's bandwidth requirements, where high values are typically observed when this mechanism is triggered.

In comparison to per-frame streaming~\cite{mossel}, we transmit data per block which allows the recovery from network outages as well as advanced streaming strategies controlled by the remote user.
Therefore, depending on the possibly very high number of eligible blocks from streaming, \eg all visible blocks after re-connection, scene updates may appear unordered and patch-by-patch which can affect the subjective latency (see the supplemental video).
However, due to the controllable strategies, the objective latency until these visible data are fully transmitted is much smaller than for inflexible frame-based approaches.

\subsection{Scene Model Completeness and Visual Quality}

\begin{figure}[t]
    \centering
    \subfigure[Hash Map by Nie{\ss}ner \etal \cite{niessner}.]
    {
        \centering
        \includegraphics[width = 0.475\columnwidth, height = 0.5\textheight, keepaspectratio]{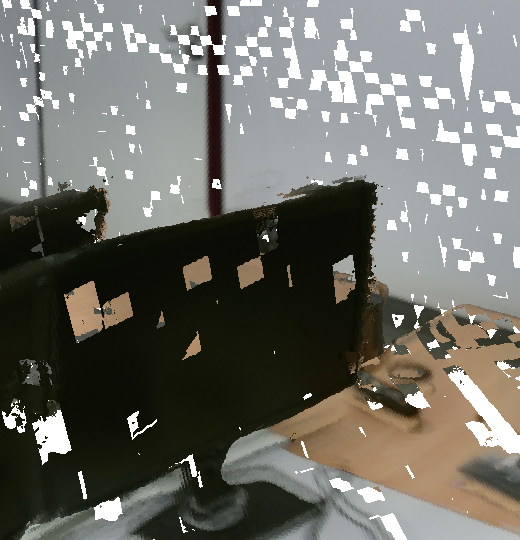}
        %\label{fig:completeness_old}
    }
    \hfill
    \subfigure[Our Hash Map Data Structure.]
    {
        \centering
        \includegraphics[width = 0.475\columnwidth, height = 0.5\textheight, keepaspectratio]{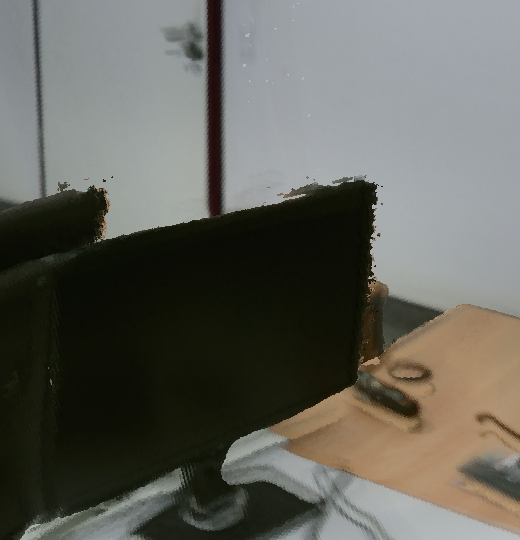}
        %\label{fig:completeness_ours}
    }
    \caption{Visual comparison of model completeness for the \emph{pool} dataset: While previous hash maps allow failures, our hash data structure ensures hole-free reconstructions during transmission to an exploration client.}
    \label{fig:completeness}
\end{figure}

In addition to the bandwidth analysis, we have also evaluated the model completeness during transmission for our novel hash map data structure in comparison to previous techniques that allow failures~\cite{niessner}.
Thus, we measured the model size in terms of voxel blocks at the reconstruction client, where the streaming starts, and at the exploration client, where the data is finally transmitted to.
To reduce side effects caused by distributing the computational load, we have chosen a package size of 1024 blocks (see \autoref{tab:timings}).
Whereas previous GPU hashing techniques work well for 3D reconstruction and failures can be cleaned up in subsequent frames, they are not suitable for large-scale collaboration scenarios where blocks are often sent only once to save bandwidth.
Insertion and removal failures will, hence, lead to holes in the reconstruction that cannot be repaired in the future (see \autoref{fig:completeness}).

\begin{figure}[t]
    \centering
    \subfigure[With Color Interpolation.]
    {
        \centering
        \includegraphics[width = 0.475\columnwidth, height = 0.5\textheight, keepaspectratio]{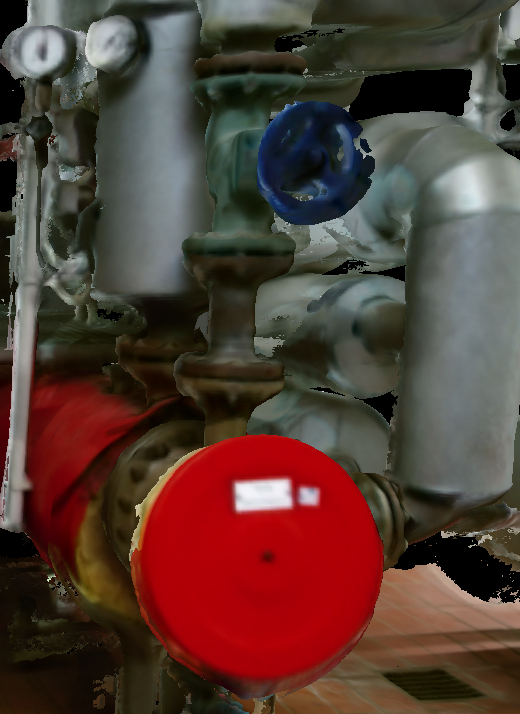}
        %\label{fig:mc_approximation_good_interp}
    }
    \hfill
    \subfigure[Without Color Interpolation.]
    {
        \centering
        \includegraphics[width = 0.475\columnwidth, height = 0.5\textheight, keepaspectratio]{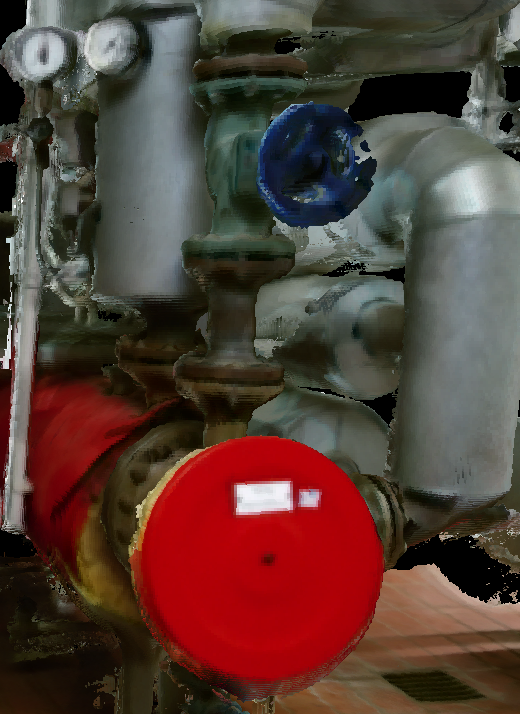}
        %\label{fig:mc_approximation_good_nointerp}
    }
    \caption{Visual comparison of our scene encoding for the \emph{heating\_room} dataset: Compared to standard mesh generation techniques that use linear interpolation, our scene encoding achieves a similar quality without interpolation in real-world scenes.}
    \label{fig:mc_approximation_good}
\end{figure}

\begin{figure}[t]
    \centering
    \subfigure[With Color Interpolation.]
    {
        \centering
        \includegraphics[width = 0.475\columnwidth, height = 0.5\textheight, keepaspectratio]{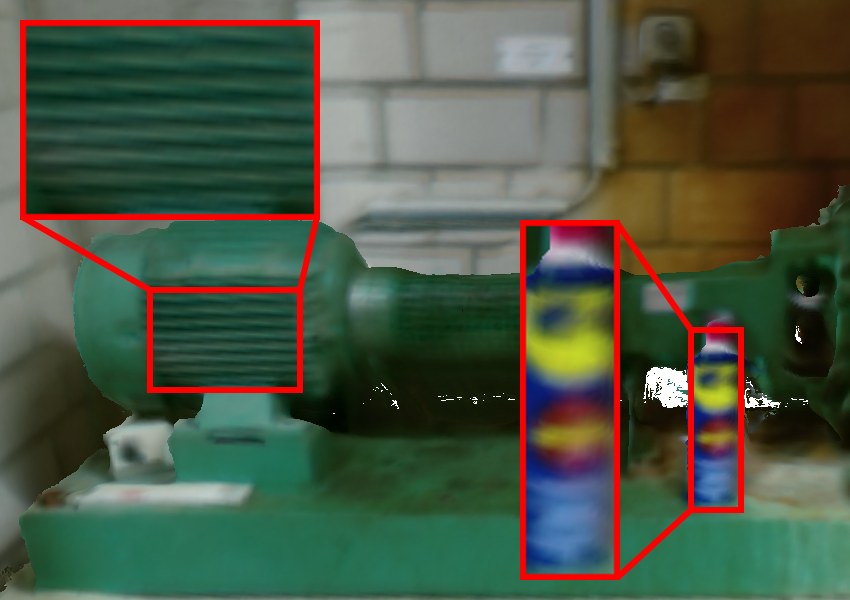}
        %\label{fig:mc_approximation_bad_interp}
    }
    \hfill
    \subfigure[With Color Interpolation.]
    {
        \centering
        \includegraphics[width = 0.475\columnwidth, height = 0.5\textheight, keepaspectratio]{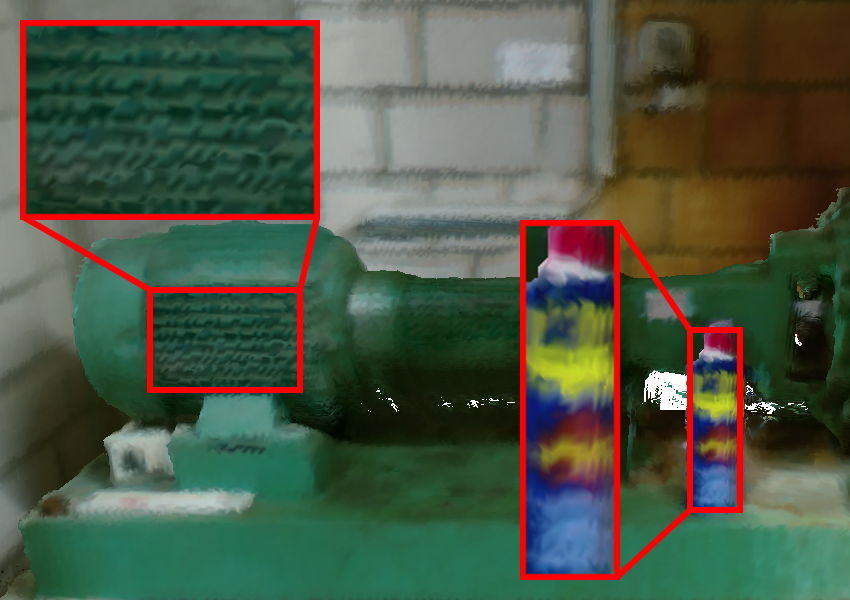}
        %\label{fig:mc_approximation_bad_nointerp}
    }
    \caption{Challenging cases: For highly textured objects and sharp edges with high contrasts, our approximation introduces small artifacts.}
    \label{fig:mc_approximation_bad}
\end{figure}

We also provide a qualitative visual comparison of our bandwidth-optimized scene representation based on Marching Cubes indices.
In order to reduce the bandwidth requirements by over 90\%, we omitted the interpolation of vertex positions and colors.
\autoref{fig:mc_approximation_good} shows a comparison between our approximation and the interpolated mesh, where both representations have been reconstructed using a voxel resolution of 5mm.
While the interpolated model has a smooth appearance, the quality of our approximation is slightly lower at edges but, otherwise, resembles the overall visual quality quite well.
However, for small highly textured objects, staircase artifacts become visible and lead to worse reconstruction results (see \autoref{fig:mc_approximation_bad}).
Note that our system allows compensating this issue by using our projective texture mapping approach to enable higher resolution information on demand.

\subsection{Live Remote Collaboration}

To verify the usability of our framework, we conducted a live remote collaboration experiment where a local user and two remotely connected users collaboratively inspect the local user's environment supported by audio-communication (\ie via Voice over IP (VoIP)).
For this experiment, we selected people who were unfamiliar to our framework and received a briefing regarding the controls.
Furthermore, these user have never been in the respective room before.

While one person took the role of a local user operating the acquisition device, two different remotely connected exploration clients provide support regarding maintenance and safety.
The exploration clients can interactively inspect the acquired scene, \ie the maintenance expert guides the person operating the acquisition device to allow the observation of measurement instruments.
By allowing scene resets, where parts of the scene can be updated on demand, our system allows certain scene manipulations such as opening the door to a switch board that has to be checked by the maintenance expert.
Furthermore, the scene model can be visualized at higher texture resolution based on the transmission of the live-captured RGB image upon request and its usage in a separate virtual 2D display or directly on the scene geometry.
This allows checking instruments or even reading text (see supplemental material for further details and evaluation).
Measurements performed based on the controllers belonging to the HMD devices are of sufficient accuracy to allow detecting safety issues or select respective components for replacement.
The interaction flow of this experiment is also showcased in the supplemental video.
In addition to the Kinect v2, we also used an ASUS Zenfone AR ($ 224 \times 172 $ pixels, up to 15Hz) for RGB-D acquisition.
However, the limited resolution and frame rate affect the reconstruction quality obtained with the smartphone.

Furthermore, the users testing our framework particularly liked the options to reset certain scene parts to get an updated scene model as well as the possibility of interacting with the scene by performing measurements and inspecting details like instrument values.
After network outages or wanted disconnections from the collaboration process, the capability of re-connecting to re-explore the in-the-meantime reconstructed parts of the scene was also highly appreciated and improved the overall experience significantly.
In fact, they reported a good spatial understanding of the environment.

\subsection{Limitations}

Despite allowing an immersive live collaboration between an arbitrary number of clients, our system still faces some limitations.
In particular, the acquisition and reconstruction of a scene with a RGB-D camera may be challenging for unexperienced users, who tend to move and turn relatively fast resulting in high angular and linear velocities as well as potential motion blur.
As a consequence, the reconstruction is more susceptible to misalignments.
Whereas loop-closure techniques~\cite{Dai:2017} compensate this issue, their uncontrollable update scheme during loop closing would cause nearly the entire model to be queued for streaming.
This would impose much higher bandwidth requirements to the client connections and prohibit remote collaboration over the Internet.
Submap approaches~\cite{Kaehler:2016} avoid this problem, but issues such as the submap jittering caused by progressive relocalization during the capturing process have to be handled carefully in order to preserve an acceptable VR experience and require a respective evaluation in the scope of a comprehensive user study.
Furthermore, we stream the virtual model in the TSDF voxel representation between the reconstruction client and the server which requires both to be in a local network.
However, the increasing thrust in cloud services could fill this gap.
While we believe that the usability of our novel system significantly benefits from mobile devices with built-in depth cameras, the current quality and especially the frame rate of the provided RGB-D data is inferior compared to the Kinect family resulting in low-quality reconstructions.

%-------------------------------------------------------------------------
\section{Conclusion}

We presented a novel large-scale 3D reconstruction and streaming framework for immersive multi-client live telepresence that is especially suited for remote collaboration and consulting scenarios.
Our framework takes RGB-D inputs acquired by a local user with commodity hardware such as smartphones or the Kinect device from which a 3D model is updated in real-time.
This model is streamed to the server which further manages and controls the streaming process to the, theoretically, arbitrary number of connected remote exploration clients.
As such as system needs to access and process the data in highly asynchronous manner, we have built our framework upon -- to the best of our knowledge -- the first thread-safe GPU hash map data structure that guarantees successful concurrent insertion, retrieval and removal on a thread level while preserving key uniqueness required by current voxel block hashing techniques.
Efficient streaming is achieved by transmitting a novel, compact representation in terms of Marching Cubes indices. In addition, the inherently limited resolution of voxel-based scene representations can be overcome with a lightweight projective texture mapping approach which enables the visualization textures at the resolution of the depth sensor of the input device.
As demonstrated by a variety of qualitative experiments, our framework is efficient regarding bandwidth requirements, and allows a high degree of immersion into the live captured environments.

%-------------------------------------------------------------------------

\acknowledgments{
This work was supported by the DFG projects KL 1142/11-1 (DFG Research Unit FOR 2535 Anticipating Human Behavior) and KL 1142/9-2 (DFG Research Unit FOR 1505 Mapping on Demand).
}

\bibliographystyle{abbrv-doi}

\bibliography{IEEEVR_2018_JOURNAL/literature}

\end{document}